\documentclass[twocolumn]{aastex631}

\usepackage{comment}
\usepackage{graphicx,floatrow,amsmath,gensymb,rotating}
\usepackage{subfigure}
\usepackage{natbib}
\usepackage{xcolor}
\usepackage{threeparttable,booktabs}

\newcommand{\lcx}[1]{\textcolor{black}{ #1}}

\newcommand{\rv}[1]{#1}

\newcommand{\nmem}{\textcolor{black}{63,827~}}
\newcommand{\ntot}{\textcolor{black}{15,940~}}
\newcommand{\nscr}{\textcolor{black}{10,499~}}
\newcommand{\nns}{\textcolor{black}{5,441~}}
\newcommand{\nnsa}{\textcolor{black}{2,083~}} 
\newcommand{\nnsc}{\textcolor{black}{3,358~}}
\newcommand{\nextr}{\textcolor{black}{17,552~}}

\newcommand{\mgii}{\ion{Mg}{2}] $\lambda2799$}
\newcommand{\civ}{\ion{C}{4} $\lambda1549$}

\newcommand{\nv}{\ion{N}{5} $\lambda1241$}

\newcommand{\heii}{\ion{He}{2} $\lambda1640$}
\newcommand{\ciii}{\ion{C}{3}] $\lambda1909$}
\newcommand{\ovi}{\ion{O}{6} $\lambda1034$}
\newcommand{\lya}{Ly$\alpha$ $\lambda1215$}
\newcommand{\ha}{H$\alpha$}
\newcommand{\hb}{H$\beta$}

\newcommand{\escma}{ergs~s$^{-1}$~cm$^{-2}$~\AA$^{-1}$}
\newcommand{\escm}{ergs~s$^{-1}$~cm$^{-2}$}

\newcommand{\kms}{km $\mathrm{s}^{-1}$} 
\newcommand{\elixer}{\texttt{ELiXer}}

\def\arcsec{$^{\prime\prime}$}
\def\arcmin{$^{\prime}$}

\begin{document}

\title{The Hobby-Eberly Telescope Dark Energy Experiment Survey (HETDEX) Active Galactic Nuclei Catalog: the Fourth Data Release}

\author[0000-0001-5561-2010]{Chenxu Liu}
\affiliation{South-Western Institute for Astronomy Research, Yunnan University, Kunming 650500, People's Republic of China}
\correspondingauthor{Chenxu Liu}
\email{cxliu@ynu.edu.cn}

\author[0000-0002-8433-8185]{Karl Gebhardt}
\affiliation{Department of Astronomy, The University of Texas at Austin, 2515 Speedway, Austin, TX 78712, USA}

\author[0000-0002-2307-0146]{Erin Mentuch Cooper}
\affiliation{Department of Astronomy, The University of Texas at Austin, 2515 Speedway, Austin, TX 78712, USA}
\affiliation{McDonald Observatory, The University of Texas at Austin, 2515 Speedway, Austin, TX 78712, USA}

\author[0000-0002-8925-9769]{Dustin Davis}
\affiliation{Department of Astronomy, The University of Texas at Austin, 2515 Speedway, Austin, TX 78712, USA}

\author[0000-0001-7240-7449]{Donald P. Schneider}
\affiliation{Department of Astronomy \& Astrophysics, The Pennsylvania State University, University Park, PA 16802, USA}
\affiliation{Institute for Gravitation and the Cosmos, The Pennsylvania State University, University Park, PA 16802, USA}

\author[0000-0001-7039-9078]{Matt J. Jarvis}
\affiliation{Astrophysics, Department of Physics, University of Oxford, Keble Road, Oxford, OX1 3RH, UK}
\affiliation{Department of Physics and Astronomy, University of the Western Cape, Robert Sobukwe Road, 7535 Bellville, Cape Town, South Africa}

\author[0000-0003-2575-0652]{Daniel J. Farrow}
\affiliation{Centre of Excellence for Data Science, Artificial 
Intelligence and Modelling, University of Hull, Cottingham Road, Hull, HU6 7RX, UK}
\affiliation{E.A. Milne Centre for Astrophysics, University of Hull, Cottingham Road, Hull, HU6 7RX, UK}

\author[0000-0001-8519-1130]{Steven L. Finkelstein}
\affiliation{Department of Astronomy, The University of Texas at Austin, 2515 Speedway, Austin, TX 78712, USA}

\author[0000-0003-2332-5505]{\'{O}scar A. Ch\'{a}vez Ortiz}
\altaffiliation{NASA FINESST Fellow}
\affiliation{Department of Astronomy, The University of Texas at Austin, 2515 Speedway, Austin, TX 78712, USA}

\author{(The HETDEX Collaboration)}

\begin{abstract}

We present the Active Galactic Nuclei (AGN) catalog from the fourth data release (HDR4) of the Hobby-Eberly Telescope Dark Energy Experiment Survey (HETDEX). HETDEX is an untargeted spectroscopic survey. HDR4 contains \lcx{345,874} Integral Field Unit (IFU) observations from January 2017 to August 2023 covering an effective area of \lcx{62.9}~deg$^2$.  With no imaging pre-selection, our spectroscopic confirmed AGN sample includes low-luminosity AGN, narrow-line AGN, and/or red AGN down to $g\sim25$. This catalog has \ntot AGN across the redshifts of $z=0.1\sim4.6$, giving a raw AGN number density of 253.4~deg$^{-2}$. Among them, \nscr (66\%) have redshifts either confirmed by line pairs or matched to the Sloan Digital Sky Survey Quasar Catalog. For the remaining \nns AGN, \nnsa are single broad line AGN candidates, while the remaining \nnsc are single intermediate broad line (full width at half maximum, FWHM $\rm \sim1200~km~s^{-1}$) AGN candidates. A total of \lcx{4,060} (39\%) of the \nscr redshift-confirmed AGN have emission-line regions $3\sigma$ more extended than the image quality which could be strong outflows blowing into the outskirts of the host galaxies or ionized intergalactic medium.
\end{abstract}

\section{Introduction} \label{sec_intro}

\lcx{The Hobby-Eberly Telescope Dark Energy Experiment Survey\footnote{\url{https://hetdex.org/}} (\citealp[HETDEX;][]{Gebhardt2021}) is a groundbreaking survey designed to map the large-scale structure of the universe and constrain the nature of dark energy at cosmic noon through the observation of Lyman Alpha Emitters (LAEs) and other extragalactic populations. However, beyond its cosmological objectives, HETDEX offers unique opportunities to probe active galactic nuclei (AGN) within its untargeted spectroscopic survey. Its integral field unit (IFU) configuration enables the untargeted spectroscopic observations with no imaging pre-selections. This allows the identification of AGN, including populations that are typically underrepresented in traditional surveys, such as the low-luminosity AGN, narrow-line AGN, and/or red quasars at $z>1$. These AGN, which are usually rejected by the imaging pre-selections in traditional surveys, can be efficiently revealed using IFU surveys \citep[e.g.][]{Jarvis2005, Gavignaud2006, Liu2022a, Liu2022c}.}

\lcx{AGN are pivotal in understanding the co-evolution of galaxies and their central supermassive black holes (SMBHs). They serve as laboratories for studying phenomena such as galaxy feedback, accretion physics, and the role of ionizing radiation in the intergalactic medium. HETDEX, with its large field of view and untargeted spectroscopic approach, is particularly suited to addressing challenges in AGN science, such as uncovering extended Lyman Alpha blobs (LABs) around AGN. The spatially resolved spectra can help understand the origins of the extended emission-line regions around AGN and quantify the potential mass exchanges between the central SMBHs and their host galaxies.} 


The second HETDEX Data Release (HDR2) AGN catalog is discussed in \cite{Liu2022a,Liu2022b}. In this paper, we release the 15,877 AGN identified from the fourth HETDEX Data Release (HDR4), which includes observations from January 2017 through August 2023. Section \ref{sec_obs} provides an overview of the HETDEX observations and data release history.
Section \ref{sec_select} introduces the identification methods of AGN detections. Since HETDEX is an IFU survey, a single AGN can be detected with multiple fiber detections. Section \ref{sec_fof} discusses how we deal with duplicate detections and generate the unique AGN catalog. \rv{Section~\ref{sec_completeness} discusses the completeness of this catalog.} Section \ref{sec_statistics} presents the statistics of the catalog (redshift and $g$-band magnitudes). Section \ref{sec_catalog} provides details of the data construction of the catalog file. Section \ref{sec_summary} summarizes this catalog release.

\section{Observations, Data Reductions, and Data Release History} \label{sec_obs}

HETDEX is an untargeted spectroscopic survey aiming at mapping the $1.88<z<3.52$ universe and measure its expansion rate to 1\% precision by providing the redshifts of a million Lyman-Alpha Emitters (LAEs) over 540~deg$^2$. The untargeted spectroscopic observations with no imaging pre-selections are acquired by the Visible Integral-field Replicable Unit Spectrograph (\citealp[VIRUS;][]{Hill2021}). VIRUS has 78 51\arcsec$\times$51\arcsec~IFUs spanning on the central 18\arcmin~field of view of the 22\arcmin~focal plane on the 10-meter Hobby-Eberly Telescope (\citealp[HET;][]{Ramsey98}). The survey began from January 2017 with 16 IFUs mounted. IFUs were installed on the telescope gradually. All IFUs were completed with their installation in August 2021. Each IFU bundle consists of 448 1.5\arcsec~diamter fibers. Upon completion, each HETDEX exposure collects 34,944 fiber spectra. 
The survey was recently completed with all observations in August 2024. The spectral coverage is from 3500~to~5500~\AA~with the resolving power of $R\sim800$. 

The survey design, data reductions, and emission-line detection algorithms are detailed in \cite{Gebhardt2021}. Each ``shot'' consists of three 6-minute dithered exposures to fill the space between fibers of individual IFU. 
This procedure yields a filling factor of 1/4.6 on the completion of all 78 IFUs. Sky subtraction is done with single amplifier (112 fibers). 
A HETDEX $g$-band reconstructed image (integrated over 4400 - 5200~\AA) is generated for each shot.  Astrometry and flux of the reconstructed images are calibrated to a collection of stars in the Sloan Digital Sky Survey (SDSS) fifteenth Data Release (DR15) \citep{York2000,Abazajian2009}, Gaia DR2 \citep{Gaia2018}, Panoramic Survey Telescope and Rapid Response System \citep[PanStarrs;][]{Chambers2016,Flewelling2020}, and United States Naval Observatory \citep{Roeser2010} catalogs. All HETDEX flux calibrated spectra in this paper are not corrected for foreground Galactic extinction. 

Image quality and its error is calculated as the mean Full Width at Half Maximum (FWHM) of the Point Spread Functions (PSFs) of bright stars in the frame (typically $\sim30$ stars). The typical image quality is {\tt FWHM\_virus~$\sim$~1.8\arcsec}. 
At the project's nominal depth, the PSF-weighted line flux sensitivity reaches a $5\sigma$ limit of $3.5 \times 10^{-17}$~ergs~cm$^{-2}$~s$^{-1}$ at $\lambda \sim 5000$~\AA.

The first HETDEX Data Release (HDR1) contains observations through June 2019 and was commissioning so kept internal. HDR2 is through June 2020. The HDR2 source catalog (200,000 sources with 51,863 $z\sim2.5$ LAEs) is available in \cite{Cooper2023}. The HDR2 AGN catalog has 5,322 AGN \citep{Liu2022a}, hereafter Liu22. The third HETDEX Data Release (HDR3) is through August 2022. The first public data release (PDR1) based on HDR3 will contain $\sim$~300 million fiber spectra and 500,000 sources (Mentuch Cooper et al., in preparation 2025). Observations of HDR4 were completed in August 2023 with the full survey 72\% completed. This paper describes the HETDEX HDR4 AGN catalog.

\section{AGN detection identification} \label{sec_select}

The most important sources that meet HETDEX's primary goal of measuring the cosmology at cosmic noon are the LAEs. 
The main source catalog is focused on the narrow emission line detections down to signal-to-noise ratio (SNR) levels of $\sim4.5\sigma$. 
The AGN whose lines are usually broad should be identified independently from the main source catalog. 

Liu22 presents details of the AGN identification process in Section~4 and summarizes it in a flowchart (Figure~5). Here, we briefly review this process. We start from the direct detection products of the HETDEX pipeline (Section~\ref{sec_pipe}), narrow down the parent sample with quality control (Section~\ref{sec_parent}), run the AGN search algorithms (Section~\ref{sec_method}), and collect the AGN detection candidates. These candidates are then visually inspected (Section~\ref{sec_visual}). 

\subsection{HETDEX SOURCE DETECTION}\label{sec_pipe}

The HETDEX project focuses on the high completeness and the low false positive rate of detecting the LAEs. Continuum bright sources produce a high false positive rate of line detections because windows of continuums sometimes can be incorrectly fit with significant broad lines during the 3-dimensional element search of line emission signals. However, broad emission lines are rare in the LAE population. Therefore, it is more efficient to separate the continuum bright sources (Section~\ref{sec_cont}) and the line emission detections (Section~\ref{sec_lines}) for different scientific purposes.

\subsubsection{The Continuum Catalog}\label{sec_cont}

We measure the counts in a 200~\AA~window in the blue (3700 - 3900~\AA) and in the red (5100 - 5300~\AA). A threshold is set to separate the continuum detections and the line emission detections. If either region of a fiber spectrum contains more than 50 counts per 2~\AA~pixel on average ($g\sim22.5$), the position of the fiber is marked as a continuum source and searched with a $15\times15$ grid raster of 0.1\arcsec~spatial bins to locate the position that has the lowest $\chi^2$ from the PSF fit. A PSF-weighted extraction spectrum is then generated for the continuum source, and no further wavelength resolved element search is performed to this continuum source.

\subsubsection{The Line Catalog}\label{sec_lines} 

All non-Continuum fiber spectra are examined for line emission signals in the 3-dimensional space of the IFU datacubes. The initial search uses a $3\times3$ grid of 0.5\arcsec~in the spatial direction and 8~\AA~element search in the wavelength direction. The large bins in the initial search resolution, for each IFU, is 448 fibers per dither $\times$ 3 dithers $\times$ 500 wavelength elements per fiber $\times$ 9 spatial elements per fiber $\sim$ 6 million element search. As of HDR4, there are 345,874 IFU observations in total, corresponding to $2\times10^{12}$ element analyses. We performed numerical simulations and determined that this initial grid search can effectively save computation time while guaranteeing the completeness to a one percent level compared to finer grids. Each resolution element is fit with a single Gaussian profile using the instrumental line width of 2~\AA. Any Gaussian line candidates (SNR $>4$, $\chi^2<3$ ) are examined with a ($5\times5$) grid of 0.15\arcsec~raster to identify the line center. Each line detection is assigned parameters from the best single Gaussian fit (wavelength, linewidth, etc) and a PSF-weighted extraction spectrum. 

\subsection{Parent Sample}\label{sec_parent}

AGN usually have strong emission lines, 
and a significant population of the AGN are continuum bright quasars. Therefore, we start from a combination of the 600,000 continuum detections (Section~\ref{sec_cont}) and the 28 million line detections (Section \ref{sec_lines}). Before running through AGN detection algorithm, we made some initial constraints on the data. First, we require that a nominal throughput, assuming a 360 s exposure time, must be greater than 0.08. We then refine the catalog by removing known issues, such as bad amplifiers, hot pixels, and meteors. Finally, we remove some artifacts, mostly bleeding cosmic ray hits, using $\chi^2_{\rm PSF}<4$ from the 2D fit. After removal of suspicious detections with the three criteria, the 29 million raw catalog is reduced to 12 million detections (500,000 continuum detections and 11.5 million line detections).

\subsection{AGN Identification Methods}\label{sec_method}

We executed \lcx{our} two AGN search methods: the Line Pair method (LP, Section \ref{sec_lp}) and the single Broad Line search (sBL, Section \ref{sec_sBL}), as detailed in Liu22. 

\subsubsection{The Line Pair Method}\label{sec_lp} 
The line pair (LP) search basically identifies emission line pairs characteristic of AGN, especially when the highly ionized lines, such as \civ, are present (SNR $>5$). This approach is applied for both the continuum catalog (Section~\ref{sec_cont}) and the line catalog (Section~\ref{sec_lines}). Considering the wavelength coverage of HETDEX (3500 - 5500 \AA), the strongest AGN emission lines visible are \ovi, \lya, \nv, \civ, \ciii, \mgii~from redshift of $z=4.32$ to $z=0.25$ (see Figure 1 in Liu22). 

For each line detection, we assume the wavelength suggested by the HETDEX pipeline's single Gaussian fit to be one of the six strongest lines and check if there are signals matched to at least another line emission. The other emission line is not limited to be one of the six lines, e.g. \heii~can be used to confirm the \civ~emission. If the full wavelength range of the line detection spectrum is detected with the strongest and the second strongest lines at $>8\sigma$ and $>5\sigma$, this object is selected as an AGN candidate with a redshift indicated by the line pair(s). \rv{Higher thresholds would result in missing real AGN, and lower thresholds would bring in too many false positives making it beyond our visual inspection ability (Section~\ref{sec_visual}). We find that with the $>8\sigma$ and $>5\sigma$ thresholds no known AGN missed and the false positive rate is controlled at $\lesssim$~50\% in our visual inspection experiments on a few small test samples. We will eventually push to lower thresholds to achieve higher completeness by allowing higher false positive rates in our final data release.} 

For a continuum detection, there is no line wavelength suggested by the HETDEX pipeline in addition to a PSF-weighted extraction spectrum. We fit a power law continuum, find the highest peak in the continuum subtracted spectrum, attempt to fit the peak as one of the six strong AGN lines, and see if any other emission line(s) match with at least another peak(s). If the spectrum of a continuum detection is detected with the two strongest emission lines at levels of $>8\sigma$ and $>5\sigma$, it is then identified as an AGN candidate. All AGN identified by the LP method are labeled with {\tt sflag$=$2} \rv{(Column 8 described in Section~\ref{sec_catalog} and provided in Table~\ref{t_catalog})}.

\subsubsection{The Single Broad Line Method}\label{sec_sBL} 
Because the wavelength coverage of HETDEX is only 2000~\AA~and the limited number of lines available in this wavelength range, many of the detections only have one line detected in their spectra. For example, at \lcx{$0.96<z<1.26$}, the only line among the six strong lines visible is \ciii. 
Another possibility is that the second strongest line in our wavelength range is not significant enough to be identified ($<5\sigma$) so the spectrum also only shows a single line (see Figure 3 in Liu22 for some examples). We fit multi-Gaussian profiles to these single lines. Our code allows two emission line components, one capturing the broad feature and the other for the narrow emission line, and at most four absorption components. If a line is detected with FWHM~$>1200$~\kms~at $>5\sigma$, the source is a broad-line AGN candidate. \rv{Again, the choice of $5\sigma$ is a result of the balance between the completeness and the false positive rate. When testing on several simulated test samples, we find that a $4.5\sigma$ threshold can marginally increase the completeness from 89\% to 91\%, while the false positive rate is dramatically increased from 50\% to 90\% compared to the $5\sigma$ threshold. $5\sigma$ is believed to be efficient in recovering real broad lines while reducing the amount of work in visual inspections.} 

Some of the single broad line (sBL) identified objects are only AGN candidates, especially the ones with intermediate broad lines FWHM~$\sim1200$~\kms; they can be AGN with intermediate massive black holes (IMBHs), or type 1.5 - 2 Seyferts whose narrow line components are as strong or stronger than their broad line components. They can also be star-bursting galaxies or galaxies undergoing other activities that can increase the gas velocity dispersion. These single intermediate broad lines (sIBLs) are labeled with {\tt agn\_flag$=$0} \rv{(Column 7 described in Section~\ref{sec_catalog} and provided in Table~\ref{t_catalog})} so that they can be easily removed from statistics.

For any sBL identified AGN candidate ({\tt sflag$=$1}\rv{, Column 8 described in Section~\ref{sec_catalog} and provided in Table~\ref{t_catalog}}), an estimated redshift (flagged by {\tt zflag$=$0}\rv{, Column 6 described in Section~\ref{sec_catalog} and provided in Table~\ref{t_catalog}}) is provided based on the wavelength of the line detection and the line emission guess. The line is mainly identified as a specific emission according to their equivalent widths (EWs). Liu22 shows in Figure 14, Figure 18, and Table 2 that for the HETDEX AGN different line emissions have different EW distributions. Within the HETDEX wavelength range, the most likely broad emission lines are (\lya$+$\nv), \civ, \ciii, \mgii, and \ovi~in a decreasing EW order. The (\lya$+$\nv) emission is usually easily distinguished from other lines because of the high EW, and the line profiles of two blended strong emissions. \ovi~usually differs from the other emissions in their noisy line shape features due to the heavy \lya~forest absorption lines. It is usually difficult to distinguish single broad \civ, \ciii, and \mgii~simply based on their EWs. We check if the stronger lines expected in the wavelength range \lcx{are} present or not to narrow down the possible line solutions. For example, assuming a single broad line at 5000~\AA~as \civ, the \lya~emission should be expected at 3921~\AA. If no line emission detected at 3921~\AA, the single broad line at 5000~\AA~is then rejected to be \civ. We checked the reliability of our redshift estimates by cross matching the sBL identified AGN with SDSS DR16Q \citep{Lyke2020}. About 90\% of our redshift estimates are correct.

\subsubsection{Cross match with SDSS DR16Q}\label{sec_sdss} 

Besides the two major AGN identification methods, the LP method described in Section \ref{sec_lp} and the sBL method detailed in Section \ref{sec_sBL}, we further cross match our parent sample (Section \ref{sec_parent}) with SDSS QSOs \rv{using the SDSS's 3\arcsec~fiber size as the matching radius}. The HDR2 AGN catalog (Liu22) used SDSS DR14Q \citep{Paris2018}. In this paper, the HDR4 AGN catalog is matched to SDSS DR16Q \citep{Lyke2020}.

A significant benefit of the cross matches is to confirm or correct the redshifts for AGN identified by the sBL method. If a guessed redshift is either confirmed or corrected by an SDSS quasar, the redshift flag is reset from {\tt zflag$=$0} to {\tt zflag$=$1} \rv{(Column 6 in Table~\ref{t_catalog})}.

Besides redshift confirmations, the cross match with SDSS quasars can also identify some missed AGN (see Figure 4 in Liu22 for one example). For quasars with spatially extended emission-line region whose PSF like continuum sources fall just outside our IFU edges, the edge fibers sometimes can catch their weak diffuse emission line signals. For these cases, their spatially resolved information is important; while the lines are too weak and/or noisy to be identified by the LP method or the sBL method. Cross matching with SDSS can help recover these AGN. 
Additionally, at higher redshifts ($z>4.32$), even \ovi~is beyond our wavelength coverage. The cross match can pick up the continuum spectra bluer than \ovi~for these high-$z$ AGN (see the first spectrum in Figure \ref{f_specs} for one example). For broad line AGN at low redshifts $z\gtrsim0.1$, their broad \hb~and \ha~lines lie beyond our red wavelength limit, the cross match can also recover these AGN (see the last spectrum in Figure \ref{f_specs} for one example).

\subsection{Visual Identification}\label{sec_visual}

Once the AGN detection candidates are identified from the parent sample as detailed in Section \ref{sec_method}, visual inspections are made of each candidate. This step can remove artifacts, such as bleeding cosmic rays misidentified by the sBL search. Continuum features of stars or galaxies wrongly identified as broad lines by the sBL search can also be removed. There are some line pairs whose weaker lines are not significant enough to be detected by the LP search ($\sim4\sigma$) and their redshift estimates are given by the remaining single broad line detections. During the visual inspections, their $\sim3\sigma$~-~$4\sigma$ emission lines can provide additional information leading to better redshift guesses. The visual inspections reduce the 130K AGN detection candidates into 63,348 visually confirmed AGN detections. 

\section{Duplicate Removal}\label{sec_fof}

\begin{figure*}[htbp]
\centering
\includegraphics[width=0.99\textwidth]{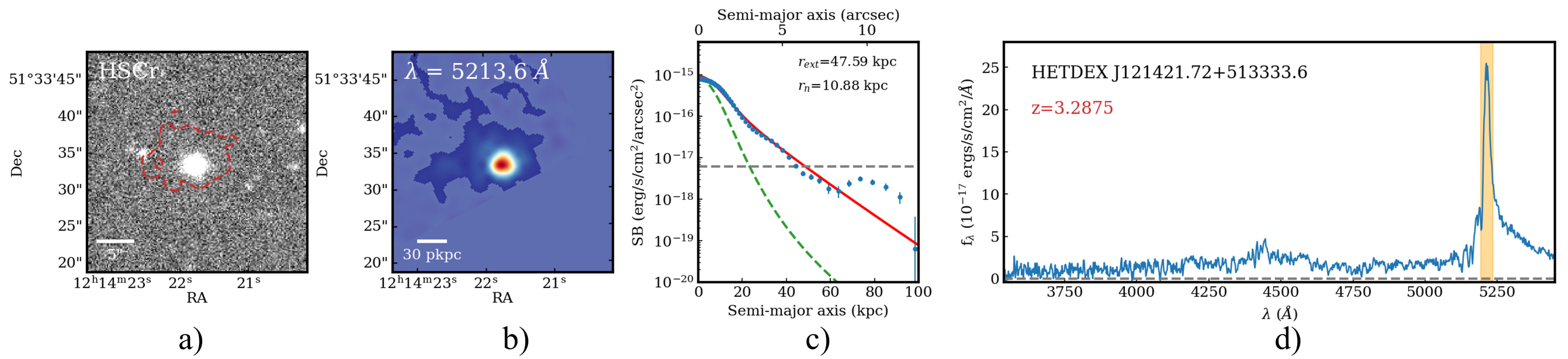}
\caption{An example of an AGN ({\tt agnid$=$407}) 
detected with extended diffuse ionized gas. a) The HSC $r$-band 30\arcsec$\times$30\arcsec~image with the red \lya~narrow-band $2\sigma$ density contour. b) The HETDEX \lya~narrow-band image. c) The radial profile of the \lya~flux density map. Blue data points are surface brightness variations. The red curve is an (PSF $+$ exponential) fit to the blue points giving a $r_{\rm ext}=47.6$~kpc. The green dashed curve gives the PSF model using the image quality of this observation. The gray dashed horizontal line marks the standard deviation in the continuum subtracted background in the line flux map. d) The spectrum extracted at the labeled coordinate in the observed frame. The yellow shaded region highlights the $\pm21$~\AA~wavelength range making the \lya~density map. We refer the readers to Mentuch Cooper et al. in preparation for more detailed analyses on the extended diffuse ionized gas around LAEs.\\}
\label{f_extend}
\end{figure*}

\begin{figure}[htbp]
\centering
\includegraphics[width=\textwidth]{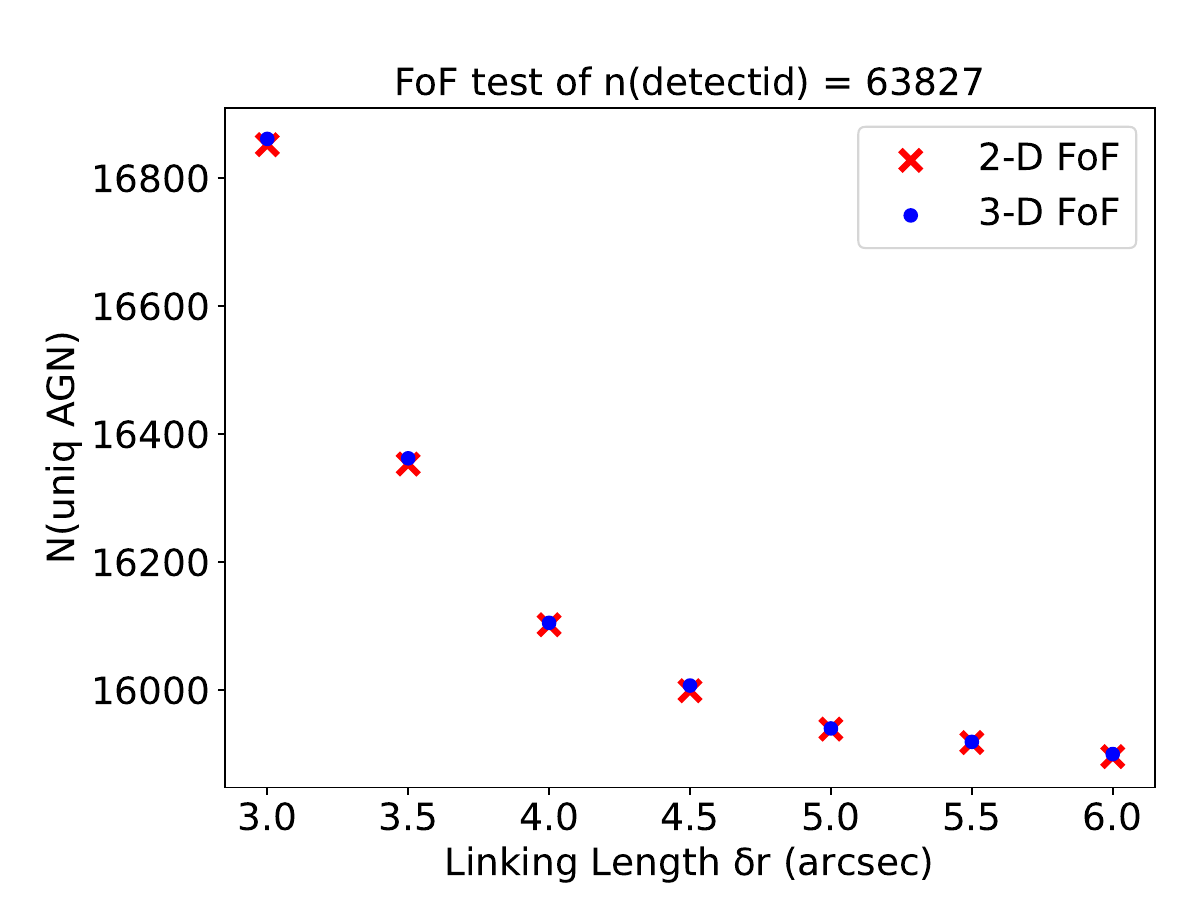}\\
\caption{The FoF experiments with linking lengths of $\delta r=$ 3.0\arcsec, 3.5\arcsec, 4.0\arcsec, 4.5\arcsec, 5.0\arcsec, 5.5\arcsec, 6.0\arcsec~for the \lcx{63,827} AGN detections. The 2-D FoF tests are the red crosses. The 3-D FoF tests ($\delta z=0.1$) are the blue data points. }
\label{f_fof}
\end{figure}

\begin{figure*}[htbp]
\centering
\includegraphics[width=0.45\textwidth]{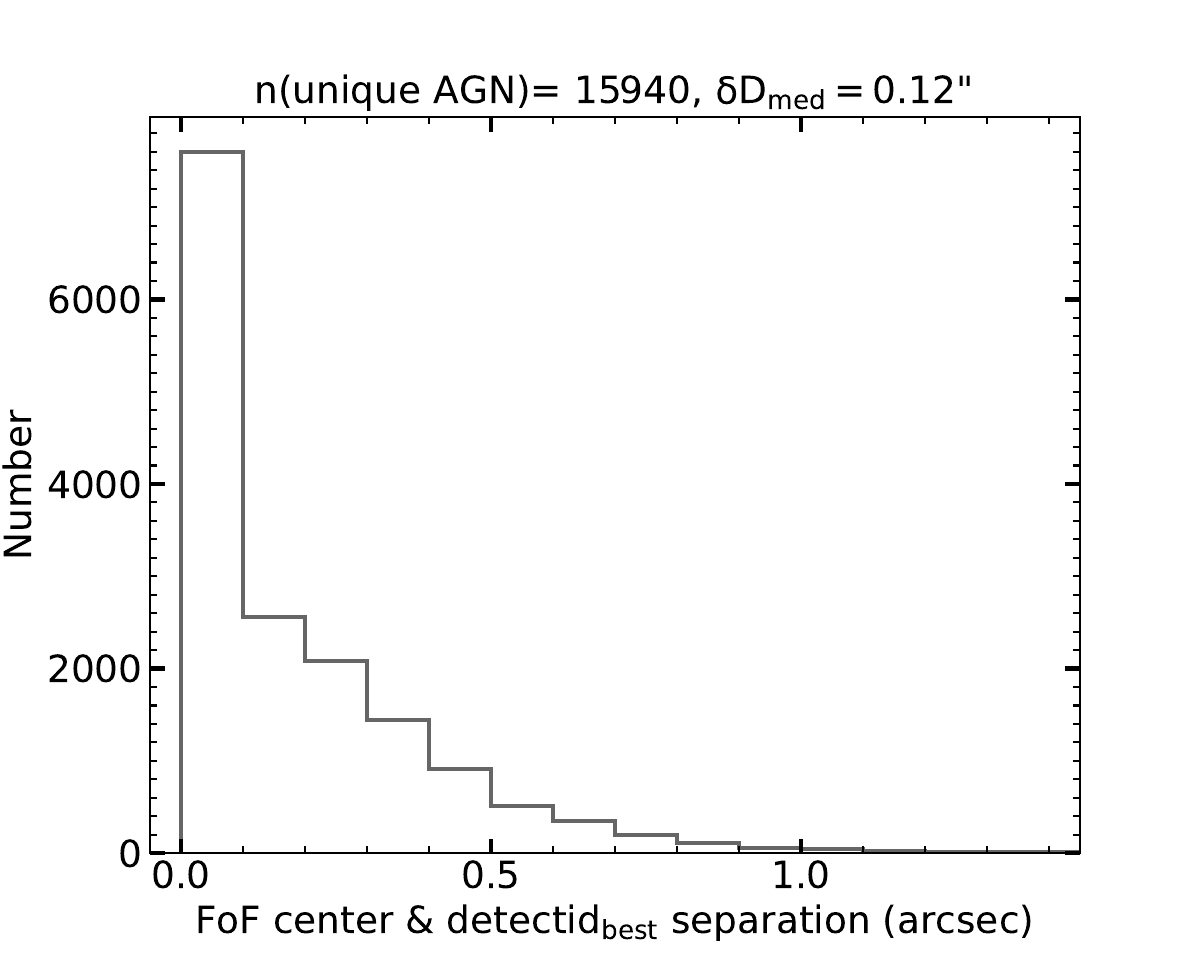}
\includegraphics[width=0.45\textwidth]{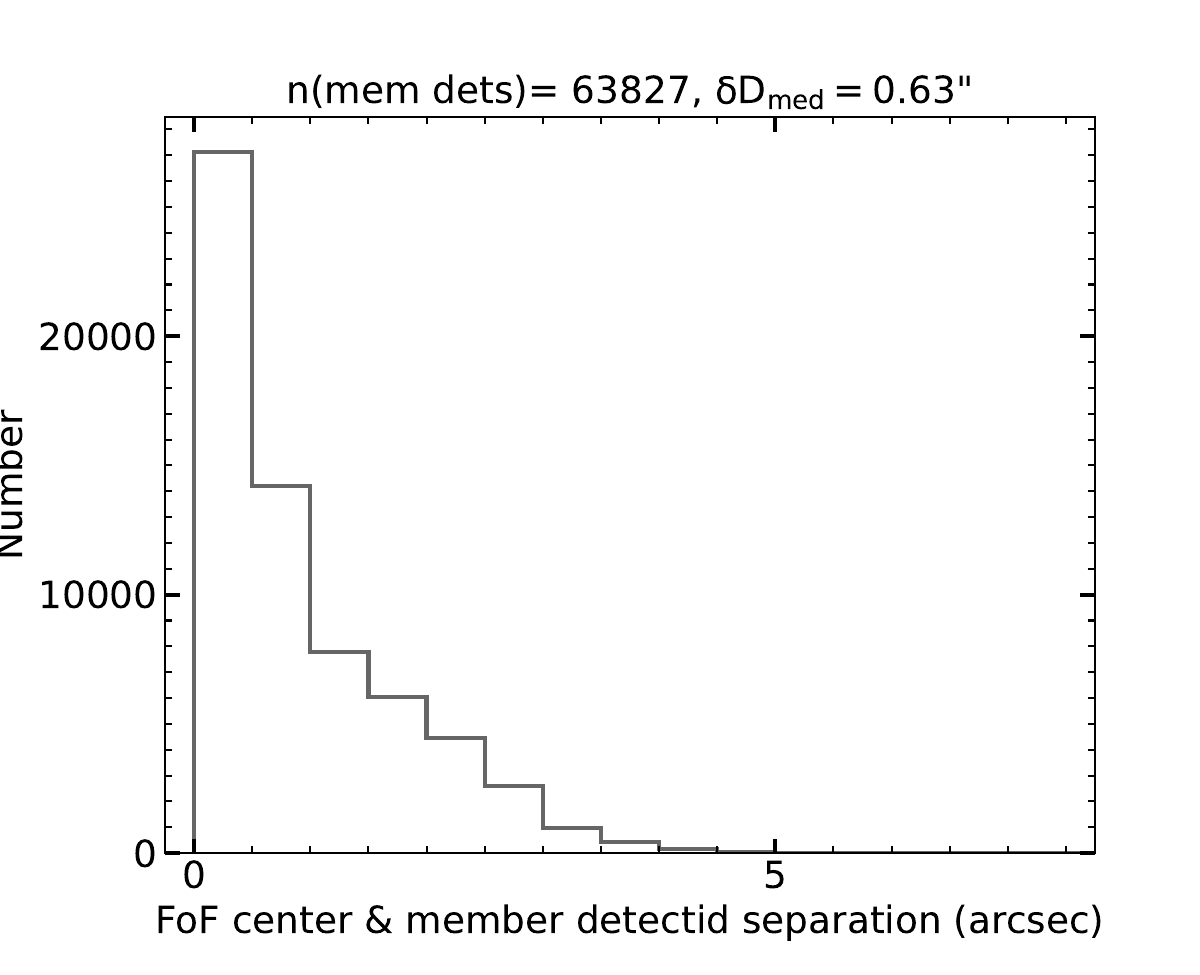}
\caption{{\it Left:} Distributions of the separations between the ``best'' detections ({\tt detectid\_best}) to their emission-line flux weighted FoF centers. {\it Right:} Distributions of the separations between all AGN detections to their emission-line flux weighted FoF centers.\\}
\label{f_roff}
\end{figure*}

As an IFU survey, a spatially extended source can illuminate multiple fibers and produce multiple detections in the HETDEX survey. This frequently occurs for the low-redshift big 
[\ion{O}{2}] galaxies. At $z\sim2-3$, about 5\% of the HETDEX LAE sample are ``\lya~blobs'' which are significantly more extended than the single PSF model (Mentuch Cooper et al. in preparation). There are also many duplicates in the \nmem AGN detections identified by the processes in Section \ref{sec_method} and Section \ref{sec_visual}. We apply friend-of-friend grouping (FoF\footnote{\url{https://github.com/HETDEX/hetdex_api/blob/master/hetdex_tools/fof_kdtree.py}}) to remove duplicates from the sample. The challenge is the choice of the linking length to make the unique AGN sample: A large one would group real AGN overdensities into a single AGN, while an insufficient linking length would leave the sample with many duplicates. During our visual inspections, many of the HETDEX AGN are found to be surrounded by diffuse ionized gas whose associated emission lines are more significantly extended than the typical FWHM$=$1.8\arcsec~PSF \citep[e.g.][]{Liu2022c}. Figure~\ref{f_extend} displays an extended AGN in our catalog. The PSF plus exponential fit ($r_{\rm ext}=47.59$~kpc) can better reproduce the \lya~density map than the PSF model alone ($r_{\rm n}=10.88$~kpc). 

There is no universal linking length suitable for the full AGN catalog. We carried out a set of FoF experiments with different linking lengths ($\delta r=$ 3.0\arcsec, 3.5\arcsec, 4.0\arcsec, 4.5\arcsec, 5.0\arcsec, 5.5\arcsec, 6.0\arcsec) to determine the optimal choice. 
Liu22 demonstrates in Figure 16 that the broad-line AGN fraction is $\sim90\%$ of the AGN in the luminosity and redshift range we probe. Considering this high fraction of broad lines and the complex line profiles in the \lya~regime, the 3-D FoF is performed with $\delta z=0.1$ in the redshift space. We also carried out 2-D FoF experiments for comparison reasons. Figure~\ref{f_fof} displays that the number of unique AGN decreases significantly as $\delta r$ increases at $\delta r\lesssim$~5\arcsec, reaching roughly a constant value at $\rm n(uniq~AGN)\sim16,000$ at $\delta r\gtrsim$~5\arcsec. We randomly examine the line flux maps of some newly added distinct AGN in the $(\delta r,\delta z)=$(4.5\arcsec, 0.1) set in addition to the $(\delta r,\delta z)=$~(5.0\arcsec, 0.1) set. 
The newly added distinct AGN in the $(\delta r,\delta z)=$(4.5\arcsec, 0.1) set are just the extended emission of existing AGN in the $(\delta r,\delta z)=$~(5.0\arcsec, 0.1) set. This suggests that $\delta r=4.5$\arcsec~is not sufficient to remove duplicates.
To avoid using a too large $\delta r$ and group real AGN pairs or groups into single ones, we decided to adopt $\delta r$ = 5\arcsec. The 2-D FoF ($\delta r~=$~5.0\arcsec)~yields the unique AGN number as $\rm n(uniq~AGN)=\lcx{15,939}$, while the 3-D FoF $(\delta r,\delta z)=$~(5.0\arcsec, 0.1) gives $\rm n(uniq~AGN)=\lcx{15,940}$. The one additional AGN in the 3-D FoF experiment is a real superposition of two separate AGN within 5\arcsec. Our final choice of FoF parameter set to group the \nmem AGN detections into a \ntot unique AGN catalog is $(\delta r,\delta z)=$~(5.0\arcsec, 0.1). Among the \ntot unique AGN sample, \nscr of them have confirmed redshifts.

The coordinate of each unique AGN is the emission-line flux weighted center of all member detections in the FoF grouping. The member detection that is closest to the FoF center of a given AGN is then the ``best'' detection and marked as {\tt detectid\_best} \rv{(Column 12 in Table~\ref{t_catalog})}. Figure~\ref{f_roff} compares the separations between {\tt detectid\_best} (left) and their emission-line flux weighted FoF centers \rv{(Column 2 \& 3 in Table~\ref{t_catalog})} to those of all AGN detections to their FoF centers (right). The median separations of the \ntot {\tt detectid\_best} to their FoF center is $\delta D=$ 0.12\arcsec, while the median separations of all AGN detections to their FoF center is $\delta D=$ 0.63\arcsec. There are many AGN with member detections larger than 3\arcsec~from their FoF centers. {\tt agnid=407} (Figure~\ref{f_extend}) even has a $>5\sigma$ member detection 8.2\arcsec~away from its FoF center. With the HETDEX resolving power (2~\AA~in the wavelength space, and 0.15\arcsec~in the fiber space, Section~\ref{sec_pipe}), {\tt agnid=407} shows no evidence of pair(s) or group(s) in the \lya~flux density map and the 1-D spectrum. 

There are \lcx{4,060} (39\%) out of the \nscr redshift-confirmed HETDEX HDR4 AGN with member detections ($>5\sigma$) more extended than their relevant PSF models {\tt (FWHM\_virus $+$ 3~FWHM\_virus\_err)}/2. Detailed comparisons between single PSF model fit and two components fit of the line flux maps can be found in Mentuch Cooper et al. in preparation. \lcx{This high fraction of AGN with extended emission-line regions could be AGN experiencing strong outflows blowing to the circum-galactic medium or even the inter-galactic medium. Alternatively, the extended emission-line regions could also be gaseous haloes lit up by the central AGN. Detailed 2-dimensional analyses on the line shifts and the line widths would help understand the physics behind this (Liu et al. in preparation). }

\section{Completeness and Contamination} \label{sec_completeness}

\begin{table*}[htbp]
\centering

\begin{tabular}{c|c|c|c|c|c|c}
\hline\hline
Field Name  &  Field center  & Area    & N$_{\rm AGN}$ & n$_{\rm AGN}$ & N$_{\rm AGN,secure\ z}$ & n$_{\rm AGN,secure\ z}$\\
           &    J2000, deg                   & deg$^2$ &        & deg$^{-2}$&        & deg$^{-2}$\\\hline
DEX-spring &  (201.7620, 52.2367) & 36.313  &  9,733  & 268.0 &  6,288 & 173.2 \\
 DEX-fall  &  ( 23.0755,  0.3056) & 20.655  &  4,819  & 233.3 &  3,411 & 165.1 \\\hline
       NEP &  (270.2904, 65.9958) &  4.238  &    981  & 231.5 &    548 & 129.3 \\
   COSMOS  &  (150.1025,  2.2396) &  1.289  &    294  & 228.1 &    179 & 138.9 \\
     SSA22 &  (334.5131,  0.3211) &  0.362  &     83  & 229.3 &     59 & 163.0 \\ 
   GOODS-N &  (189.2030, 62.2461) &  0.052  &     30  & 546.9 &     14 & 269.2 \\\hline
     Total &        --            &  62.909 & 15,940  & 253.4 & 10,499 & 166.9 \\
\hline\hline
\end{tabular}
\caption{Summary of the survey areas of the HETDEX HDR4 AGN catalog}
\begin{tablenotes}[flushleft]
\scriptsize
\item Note: DEX-spring and DEX-fall fields are HETDEX survey fields.
COSMOS and GOODS-N fields are HETDEX science verification fields.
The NEP field \citep{Oscar2023} and the SSA22 field are taken for our collaborators with their own scientific purposes.
\end{tablenotes}
\label{t_survey}
\end{table*}

\begin{figure*}[htbp]
\centering
\includegraphics[width=0.9\textwidth]{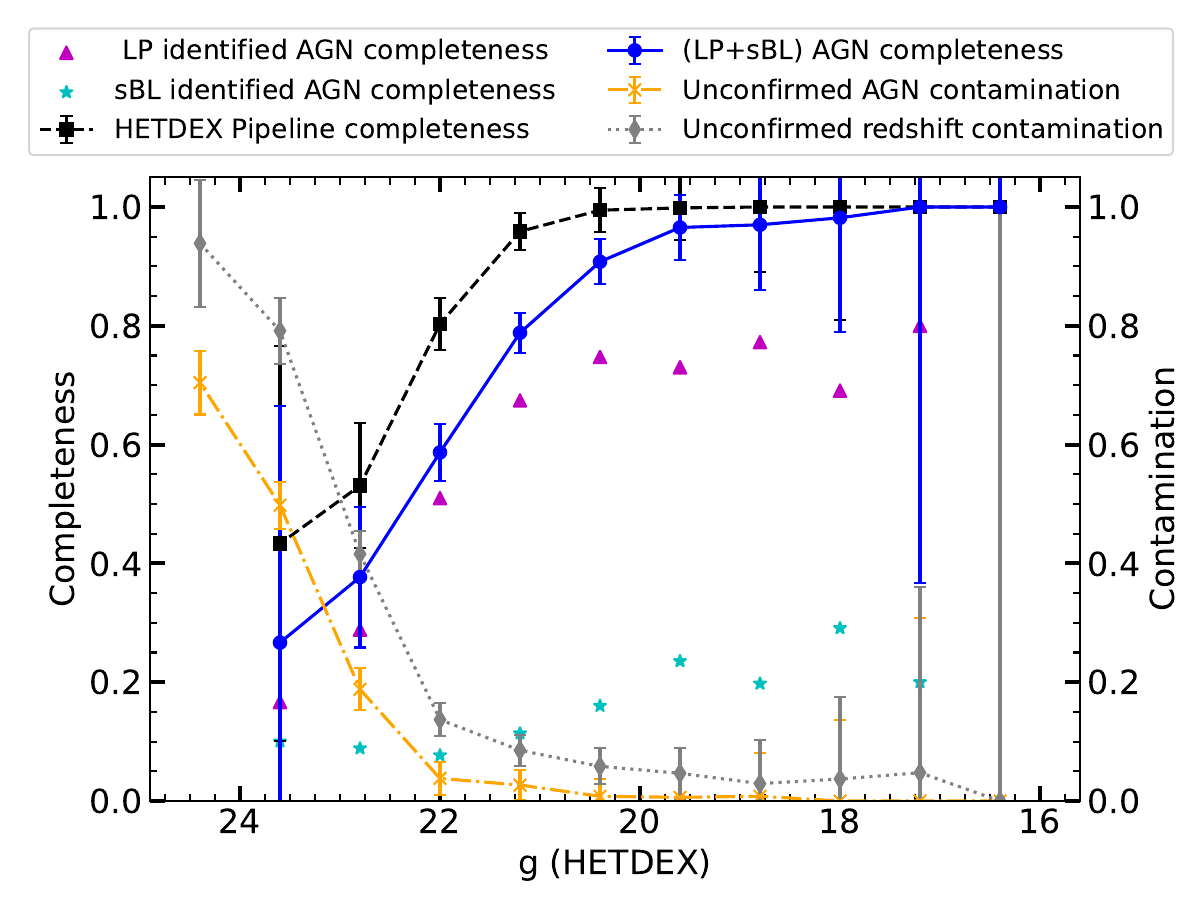}
\caption{\rv{The completeness and contamination of the HETDEX HDR4 AGN sample in each HETDEX $g$-band magnitude bin. Completeness is estimated from the cross-matched sample between the SDSS DR16Q sample and all HETDEX HDR4 fibers. The black squares connected with a dashed line show the completeness of the parent sample of this work (Section \ref{sec_pipe} and Section \ref{sec_parent}). The blue circles connected by a solid line show the completeness of the AGN identified either by the LP method (Section \ref{sec_lp}) or by the sBL method (Section \ref{sec_sBL}). The magenta triangles and the cyan stars show the completeness of the LP method (Section \ref{sec_lp}) and the sBL method (Section \ref{sec_sBL}), respectively. The gray diamonds connected by a dotted line present the contamination rate from the unconfirmed redshifts ({\tt zflag$=$0}\rv{, Column 6 in Table~\ref{t_catalog}}). The orange crosses connected by a dash-dotted line show the contamination rate from the unconfirmed AGN ({\tt agn\_flag$=$0}\rv{, Column 7 in Table~\ref{t_catalog}}).}}
\label{f_completeness}
\end{figure*}

The HETDEX HDR4 AGN catalog consists \ntot AGN after grouping duplicates into unique objects (Section~\ref{sec_fof}). Among them, \nscr AGN ({\tt zflag=1}\rv{, Column 6 in Table~\ref{t_catalog}}) have redshifts either confirmed by the LP method or the cross match with SDSS DR16Q. The remaining \nns AGN ({\tt zflag=0}) are sBL identified AGN with guessed redshifts only (see Section~\ref{sec_sBL} for details). \nnsc out of the \nns AGN are only AGN candidates with single ``intermediate'' broad lines (FWHM~$\sim$~1200~\kms) whose emission lines might be broadened by activities other than AGN, such as star bursts. They are flagged with {\tt agn\_flag=0} \rv{(Column 7 in Table~\ref{t_catalog})}. The ($\nns-\nnsc=\nnsa$) single broad line AGN combined with the \nscr {\tt zflag=1} redshift-confirmed AGN yields the \lcx{12,582} ``secure AGN'' catalog ({\tt agn\_flag=1}). 

Table \ref{t_survey} summarizes the survey area, the number of AGN, and the raw AGN density in each field in HDR4. The high AGN number density in the GOODS-N fields can either originate from real cosmic variance or from small number statistics. Even only counting the redshift secured AGN, the HETDEX AGN density is 166.9~deg$^{-2}$, which is significantly higher than that of the SDSS \rv{quasars} (56~deg$^{-2}$; \citealt{Paris2018}). This highlights the effectiveness \rv{and the high spatial completeness} in detecting \rv{optically faint} AGN with the untargeted observational strategy of HETDEX.

\rv{The HETDEX AGN identification is mostly regulated by the SNR of emission lines, as introduced in Section \ref{sec_pipe} and Section \ref{sec_method}. The detection completeness as a function of line SNR is quantitatively introduced in Figure 6 \& 7 of Liu22. Additionally, in order to calculate the HETDEX AGN luminosity function, we carefully modeled the sample completeness in \cite{Liu2022b} against all relevant parameters, including the line wavelengths, line fluxes, line widths, the relevant locations on IFUs, and redshifts for the HETDEX pipeline and the two AGN identification methods (Section~\ref{sec_lp} and Section~\ref{sec_sBL}). However, these parameters are not straightforward in comparing our completeness with other surveys. The broad-band continuum level (HETDEX $g$-band magnitude) is then estimated for each detection by multiplying the 1-D HETDEX spectrum with the SDSS $g$ filter's throughput curve using the software  Emission Line eXplorer\footnote{\url{https://github.com/HETDEX/elixer}.} (\citealp[\elixer;][]{elixer}).} 

\rv{Figure~\ref{f_completeness} presents the completeness estimation and the potential contamination of the HETDEX HDR4 AGN sample within each HETDEX $g$-band magnitude bin. We cross matched all 500 million HETDEX HDR4 fibers with the 638,022 SDSS DR16Q quasar sample (quasars flagged by redshift warnings removed) to evaluate the completeness of both the HETDEX pipeline (Section \ref{sec_pipe}) and our AGN identification methods (Section \ref{sec_method}). There are 5,872 SDSS quasars covered by HETDEX good fibers (hot fibers, fibers on the edge of the IFUs, and shots with low throughput excluded) within 1 arcsec. The weakness of using this cross-match sample to evaluate the completeness is that the sample size is limited at $g\gtrsim22.5$.} 

\rv{The recovery fraction of the SDSS DR16Q quasars brighter than $g=22.5$ is 93\% for the HETDEX pipeline (black squares) and 80\% for our AGN sample identified either by the LP method (Section \ref{sec_lp}) or by the sBL method (Section \ref{sec_sBL}) (blue circles). 
Considering all 5,872 matched SDSS DR16Q quasars, these two fractions are 92\% and 78\%, respectively.
The LP method (Section~\ref{sec_lp}) requires the strongest emission line to be $>8\sigma$. 
We would eventually push our current SNR thresholds to lower values in our final AGN data release so that the blue circles can match the black squares. Although the LP method suffers from the HETDEX's short wavelength coverage (3500 - 5500 \AA), it is the main contributor (65\%) of the AGN sample. The sBL method is also very important (13\%) in recovering AGN missed by the LP method. 
However, the sBL method also brings in AGN with unconfirmed redshifts ({\tt zflag$=$0}\rv{, Column 6 in Table~\ref{t_catalog}}) and even unconfirmed AGN ({\tt agn\_flag$=$0}\rv{, Column 7 in Table~\ref{t_catalog}}) into our sample as detailed in Section~\ref{sec_sBL}. }

\rv{For AGN brighter than $g=22.5$, 9\% of our AGN sample have unconfirmed redshifts, and 2\% are unconfirmed AGN. When including fainter bins at $g>22.5$, the two fractions are increased to 28\% and 15\%. We note here that these unconfirmed redshifts are not necessarily wrong. They are simply estimated based on single broad line wavelengths. Similarly, the unconfirmed AGN are not necessarily non-AGN. They can be low-luminosity Seyferts. With our current data by hand, we don't have enough evidences to make further confirmations.}

\section{Statistics} \label{sec_statistics}


\begin{figure}[htbp]
\centering
\includegraphics[width=\textwidth]{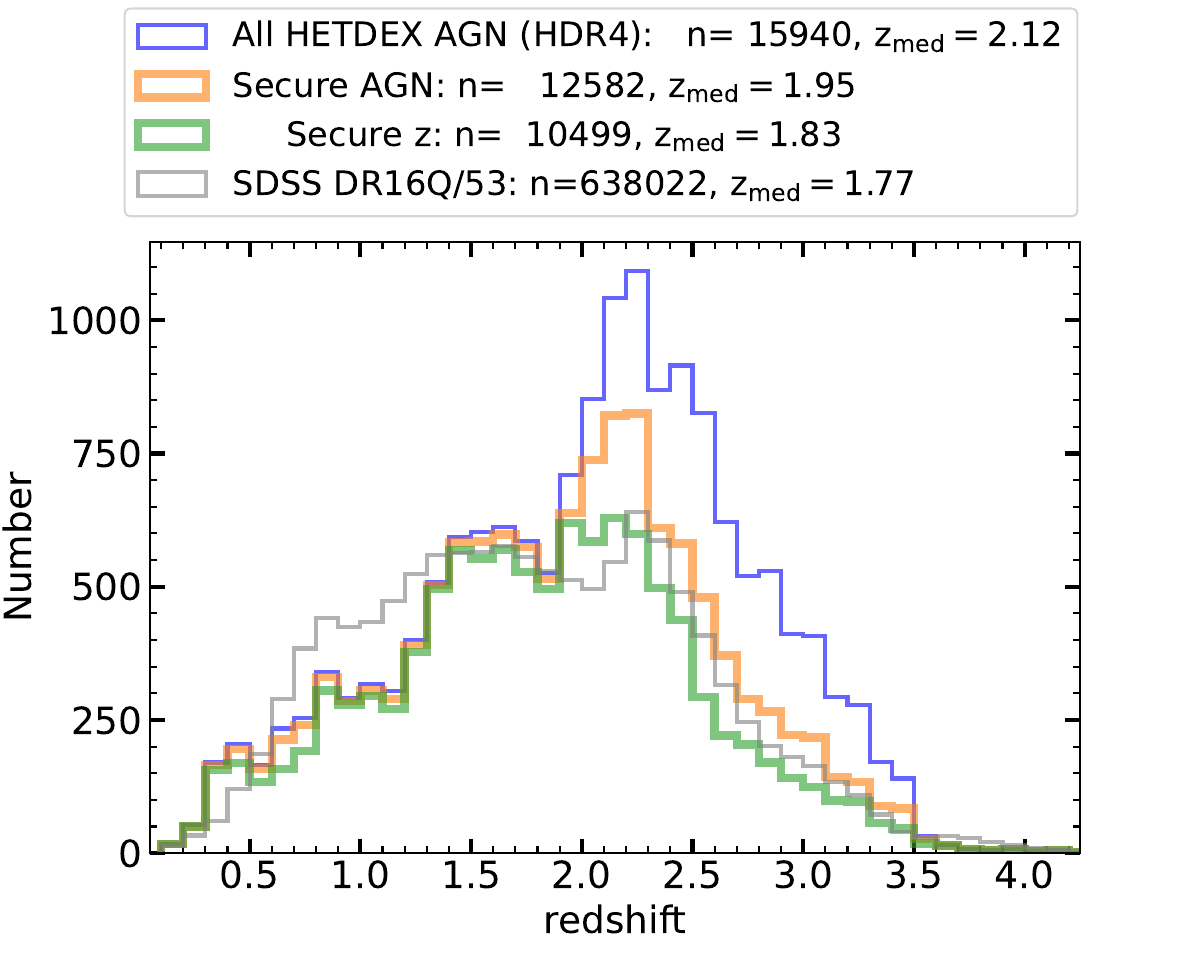}
\caption{The redshift distributions of the HETDEX HDR4 AGN catalog. The blue histogram is for the full \ntot AGN set. The orange histogram excludes the single intermediate broad line AGN candidates ({\tt agn\_flag$=$0}\rv{, Column 7 in Table~\ref{t_catalog}}). The green histogram further excludes the AGN with estimated redshifts only ({\tt zflag$=$0}\rv{, Column 6 in Table~\ref{t_catalog}}). The grey histogram is the scaled redshift distributions of the SDSS DR16Q sample \rv{to generally match the number of HETDEX AGN at $z\sim1.5$.}}
\label{f_redshift}
\end{figure}

Figure~\ref{f_redshift} shows the redshift distributions of the HETDEX HDR4 AGN sample.
The major six strong lines (\ovi, \lya, \nv, \civ, \ciii, \mgii) we used in our AGN search algorithm and the HETDEX wavelength range of 3500~-~5500~\AA~suggest a redshift range of $0.25<z<4.32$. The SDSS survey has a much broader wavelength range. By cross matching with SDSS DR16Q, a few AGN can be identified at higher ($z>4.32$) and lower ($z<0.25$) redshifts, while the majority of the HETDEX HDR4 AGN population is at $z\sim2.1$. The blue histogram is clearly more populous at $z=2\sim3.5$ than the secure AGN sub-sample ({\tt agn\_flag=1}), suggesting that many of the single ``intermediate'' broad-line AGN candidates could be non-AGN LAEs. SDSS quasars are more abundant than HETDEX AGN at $z\sim1$; this result is because it is the redshift desert of single \ciii~in the HETDEX wavelength range.


\begin{figure}[htbp]
\centering
\includegraphics[width=\textwidth]{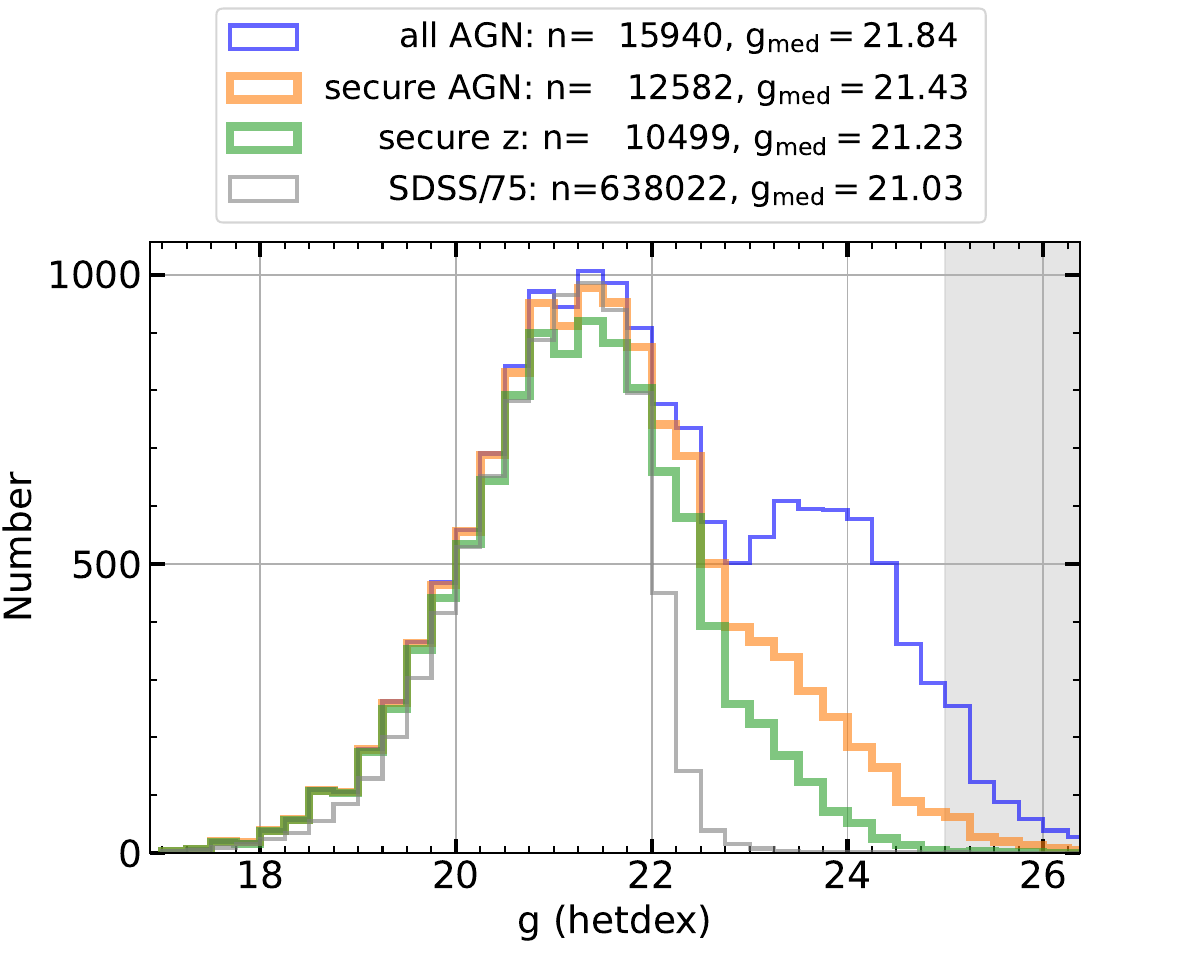}
\caption{Similar format as Figure~\ref{f_redshift} but for the HETDEX $g$-band magnitudes. While $g$-band magnitudes are plotted as computed, the $5\sigma$ depth of HETDEX is roughly 25, marked by the shaded area.}
\label{f_gmag}
\end{figure}

Figure~\ref{f_gmag} presents the $g$-band magnitude distributions of HETDEX HDR4 AGN. The typical flux density limit of HETDEX's 6-min exposures is $\sim5\times$~10$^{-19}$~\escma~($g \lesssim 25$). HETDEX is better at detecting AGN fainter than $g\sim22.5$ than SDSS. The second peak in the blue histogram at $g\sim24$ is suspicious. This peak is dominated by the {\tt agn\_flag=0} single intermediate broad line AGN candidates. It would be interesting to observe these objects in deep narrow-band images for potential emission lines at longer wavelengths in the future to determine whether they are non-AGN LAEs or low-luminosity Seyferts.

\section{The Catalog} \label{sec_catalog}


The HETDEX HDR4 AGN catalog \rv{will be made available on publication at the website}\footnote{\url{http://web.corral.tacc.utexas.edu/hetdex/HETDEX/catalogs/agn_catalog_v2.0/}} and in the online Journal (Table \ref{t_catalog}). The FITS file \citep{FITS1981} is constructed similar as the HDR2 release in Liu22 with six extensions. A simple Jupyter notebook at the public website describes how to use this FITS file in Python. The FITS extensions are as follows. 

\vspace{0.1in}
\noindent {\bf Extension 1} Table extension (\ntot rows $\times$ 56 columns): Basic information, one row for one unique AGN, arranged in
a descending order of redshifts. Column information is given in the header and summarized in Appendix~\ref{sec_append}. We explain each column with more details as follows.

\vspace{0.1in}
\noindent {\it Column 1} Sequential ID number assigned for each AGN organized in a descending redshift order from {\tt agnid} = 0 to \lcx{15,939}.

\vspace{0.1in}
\noindent {\it Column 2-3} The right ascension and declination of the emission-line flux weighted FoF center of all member detections of a given AGN (J2000). If more than one line is present within the HETDEX wavelength range, only the strongest line is used as the weight. The FoF grouping method to remove duplicates from the \nmem AGN detection sample is detailed in Section~\ref{sec_fof}.

\vspace{0.1in}
\noindent {\it Column 4-5} The best-fit redshift and its error of the 1-D spectrum of the ``best'' detection, i.e. {\tt detectid\_best}. Member detections can have slightly different redshifts due to gas kinematics.

\vspace{0.1in}
\noindent {\it Column 6} Redshift flag. Redshifts confirmed either by the LP method or matched SDSS quasars have {\tt zflag=1}, while estimated redshifts from single broad lines are flagged with {\tt zflag=0}. 

\vspace{0.1in}
\noindent {\it Column 7} AGN flag. The single ``intermediate'' broad-line AGN candidates have {\tt agn\_flag=0}; the remaining confirmed AGN have {\tt agn\_flag=1}.

\vspace{0.1in}
\noindent {\it Column 8} Source flag giving the information of a given AGN identified by which specific selection method described in Section~\ref{sec_method}. {\tt sflag=} 1, 2, and 3 indicate the AGN is identified by the sBL method (Section~\ref{sec_sBL}), LP method (Section~\ref{sec_lp}), and the cross match with SDSS DR16Q method (Section~\ref{sec_sdss}). The few {\tt sflag=0} AGN are identified by old versions of the AGN selection code.

\vspace{0.1in}
\noindent {\it Column 9} Field name in the HETDEX survey as indicated by Table~\ref{t_survey}. 

\vspace{0.1in}
\noindent {\it Column 10} The number of shots with fiber coverage at (ra,dec). A given coordinate can be observed by more than one shot. For example, some of the COSMOS fields were observed more than ten times. 

\vspace{0.1in}
\noindent {\it Column 11} Number of member detections related with a given AGN from the \nmem AGN detection sample visually confirmed by Section~\ref{sec_visual}.

\vspace{0.1in}
\noindent {\it Column 12} The member detection of a given AGN closest to its emission-line flux weighted FoF center as indicated by the coordinates recorded in Column 2 and 3. 

\vspace{0.1in}
\noindent {\it Column 13-14} The right ascension and declination of {\tt detectid\_best}. The position can be slightly off from the FoF center as the spatial detection search resolution is 0.15\arcsec~ (Section~\ref{sec_pipe}). 

\vspace{0.1in}
\noindent {\it Column 15} Offset between the coordinate of {\tt detectid\_best} ({\tt ra\_best} and {\tt dec\_best} in Column 13 and 14) and the FoF center ({\tt ra} and  {\tt dec} in Column 2 and 3). Distributions of the offset ({\tt roff}) are presented in the left panel of Figure~\ref{f_roff}.

\vspace{0.1in}
\noindent {\it Column 16-17} The HETDEX $g$-band magnitude and its 68\% confidence interval calculated by multiplying the 1-D spectrum of {\tt detectid\_best} with the SDSS $g$ filter's throughput curve using \elixer. 

\vspace{0.1in}
\noindent {\it Column 18} The identification of the fiber closest to {\tt detectid\_best}. It is a standard string with the format of {\tt yyyymmddsss\_multi\_bbb\_ccc\_ddd\_aa\_fff}. The first eight letters give the date {\tt detectid\_best} is observed. The next three digits are the shot number on the date catches {\tt detectid\_best}. {\tt bbb, ccc}, and {\tt ddd} are all three digit numbers giving the spectrograph number, the IFU number, and the IFU slot number of {\tt detectid\_best}. There are four amplifiers for each IFU: LL, LU, RL, and RU. {\tt aa} is a two letter code among the four indicating the amplifier for {\tt detectid\_best}. Each amplifier has 112 fibers. {\tt fff} is the three digit number of the specific fiber on the amplifier contributes the most flux to the 1-D PSF weighted spectrum of {\tt detectid\_best}. The full string of {\tt fiberid} can identify the raw data of a given AGN.

\vspace{0.1in}
\noindent {\it Column 19} The ID of the shot for {\tt detectid\_best}. This is an 11-digit number consistent with the first 11 digits of {\tt fiberid}.

\vspace{0.1in}
\noindent {\it Column 20} The specific amplifier for {\tt detectid\_best}. This is a 20-letter string consistent with the 13-32 letters of {\tt fiberid}.

\vspace{0.1in}
\noindent {\it Column 21} The throughput of {\tt shotid}. This parameter is described in \cite{Gebhardt2021}. Basically, it is defined as the nominal throughput treating each exposure as if it were 360~s long through a 50~m$^2$ clear aperture. Throughput varies at different wavelengths. The typical throughput at 4540~\AA~of our AGN sample is 0.15. In this paper, we require the throughput should be greater than 0.08 for an observation to be included in the catalog.

\vspace{0.1in}
\noindent {\it Column 22-23} Image quality and its error. Image quality is calculated as the mean FWHM of the Point Spread Functions (PSFs) of bright stars in the full frame (typically $\sim$~30 stars). Details can be found in \cite{Gebhardt2021} Section 6.15. The typical image quality is {\tt FWHM\_virus~$\sim$~1.8\arcsec}. 

\vspace{0.1in}
\noindent {\it Column 24} The aperture correction of {\tt detectid\_best}. Typical value of our AGN catalog is 0.92. A value of 1 means no correction. 

\vspace{0.1in}
\noindent {\it Column 25-26} The best-fit observed line wavelength and its error from \elixer's single Gaussian fit. For line detections (Section~\ref{sec_lines}), \elixer~fits to the wavelength identified by the HETDEX line detection algorithm. For continuum detections (Section~\ref{sec_cont}), \elixer~finds the most possible line to fit.

\vspace{0.1in}
\noindent {\it Column 27-46} The best-fit continuum subtracted rest-frame line fluxes, SNRs, FWHMs, and EWs of \lya, \civ, \ciii, \mgii, and \ovi. Parameters of lines falling outside the wavelength range are set to a default value of -99.0.

\vspace{0.1in}
\noindent {\it Column 47-52} The best single Gaussian fit parameters from the HETDEX's pipeline \rv{\citep{Gebhardt2021}} in the observed frame. For detections in the continuum catalog, these parameters are set to 0.

\vspace{0.1in}
\noindent {\it Column 53} The $\chi^2$ of fit to the 2-D spectrum. A high value ({\tt chi2fib\_pipe~$>$~4}) usually indicates issues such as hot pixels, bleeding cosmic rays, etc.

\vspace{0.1in}
\noindent {\it Column 54-56} Match results with the SDSS DR16Q sample. The match radius is 3\arcsec. AGN with no matches within 3\arcsec~are set with default values of {\tt rsep\_dr16q=999.0\arcsec}, {\tt z\_dr16q=-999.0}, and {\tt sdssid\_dr16q}=``?''.

\vspace{0.1in}
\noindent {\bf Extension 2} Image extension (\ntot rows $\times$ 1036 columns): 1-D spectra for the $\rm detectid_{best}$ of each AGN (no extinction correction applied). Rows are arranged in the same order as Extension 1. The wavelength array starts from 3470\,\AA\ and contains 1036 elements with a stepsize of 2.0\,\AA\null. The flux densities are in units of  $10^{-17}$ erg\,cm$^{-2}$\,s$^{-1}$\,\AA$^{-1}$. The wavelength solution and the flux density units are recorded in the extension header.

\vspace{0.1in}
\noindent {\bf Extension 3} Image extension (\ntot rows $\times$ 1036 columns): The error array for Extension 2, with the same wavelength solution and flux units.

\vspace{0.1in}
\noindent {\bf Extension 4} Table extension (\nextr rows $\times$ 4 columns): Repeat information for each AGN. Some AGN were observed multiple times. In this table, each unique observation is listed with their \texttt{shotid}. Column 1 is the \texttt{agnid}, Column 2 gives the number of repeat observations of \texttt{agnid}, and Column 3 gives the \texttt{shotid} of the observation. An AGN observed multiple times will have multiple entries in this table with the same \texttt{agnid} and \texttt{nshots}, but different \texttt{shotid}. There are \lcx{14,306} AGN observed only once with no repeats, \lcx{1,125} AGN observed twice, and \lcx{240} AGN observed more than twice. Note: \lcx{15,671} AGN can be successfully extracted with spectra at their FoF grouping center (ra,dec) in Extension 1. \lcx{269} AGN are lack of their extracted spectra because their flux weighted FoF center falling on bad amplifiers (low throughput or other issues).
     
\vspace{0.1in}
\noindent {\bf Extension 5} Image extension (\nextr rows $\times$ 1036 columns): 1-D PSF weighted extracted spectra at the FoF center, repeat observations included, no extinction correction applied. Rows are arranged in the same order as Extension 4. The wavelength solution and flux units are the same as Extension 2, and can be found in the header. 

\vspace{0.1in}
\noindent {\bf Extension 6} Image extension (\nextr rows $\times$ 1036 columns):  Error array for Extension 5.

\section{Summary} \label{sec_summary}

The HETDEX HDR4 AGN catalog has \ntot AGN. A total \nscr have redshifts either confirmed by line pairs or matched SDSS quasars, while the rest are single broad-line AGN candidates provided with redshift estimates. The catalog contains observations from January 2017 through August 2023. The effective survey area is 62.9 deg$^2$, giving a raw AGN density of 253.4 deg$^{-2}$. The redshifts range from 0.1 to 4.6, and peak at $z\sim2.1$. The $g$-band magnitudes of the faint AGN can reach the HETDEX continuum detection limit at $g\sim$ 25~mag. Approximately 39\% of the \nscr redshift-confirmed AGN sample have extended emission-line regions with $>5\sigma$ signals detected at distances significantly larger than the image quality $>$ {\tt (FWHM\_virus $+$ 3~FWHM\_virus\_err)}/2.

\vspace{0.2in}
\noindent {\bf Acknowledgments:}

\begin{acknowledgments}

CXL acknowledges supports from the ``Science \& Technology Champion Project" (202005AB160002) and the ``Top Team Project" (202305AT350002), both funded by the ``Yunnan Revitalization" (202401AT070489).

HETDEX is led by the University of Texas at Austin McDonald Observatory and Department of Astronomy with participation from the Ludwig-Maximilians-Universit\"at M\"unchen, Max-Planck-Institut f\"ur Extraterrestrische Physik (MPE), Leibniz-Institut f\"ur Astrophysik Potsdam (AIP), Texas A\&M University, The Pennsylvania State University, Institut f\"ur Astrophysik G\"ottingen, The University of Oxford, Max-Planck-Institut f\"ur Astrophysik (MPA), The University of Tokyo, and Missouri University of Science and Technology. In addition to Institutional support, HETDEX is funded by the National Science Foundation (grant AST-0926815), the State of Texas, the US Air Force (AFRL FA9451-04-2-0355), and generous support from private individuals and foundations.

The Hobby-Eberly Telescope (HET) is a joint project of the University of Texas at Austin, the Pennsylvania State University, Ludwig-Maximilians-Universit\"at M\"unchen, and Georg-August-Universit\"at G\"ottingen. The HET is named in honor of its principal benefactors, William P. Hobby and Robert E. Eberly.

The authors acknowledge the Texas Advanced Computing Center (TACC, \url{http://www.tacc.utexas.edu}) at The University of Texas at Austin for providing high performance computing, visualization, and storage resources that have contributed to the research results reported within this paper. 
\end{acknowledgments}

\clearpage
\newpage
\appendix

\section{Catalog format and column information}\label{sec_append}

The HETDEX HDR4 AGN catalog is made available in the online Journal in FITS format, and is described in Table \ref{t_catalog}. 
Other extensions to the FITS file are described in Section \ref{sec_catalog}.

\begin{longtable}{lllll} 
   \caption{Columns of Extension 1 in the FITS file.} \\  \toprule

Column & Name & dtype & Unit &      Description \\\hline
1 & agnid & int32 & - &	Sequencial number for each unique AGN \\
2 & ra & float32 & deg &	RA of the AGN (center of flux weighted FoF, J2000) \\
3 & dec & float32 & deg &	DEC of the AGN (center of flux weighted FoF, J2000) \\
4 & z & float32 & - &	Redshift \\
5 & z\_err & float32 & - &	Redshift error\\
6 & zflag & int32 & - &	zflag=0/1: 1 confirmed z; 0 estimated z \\
7 & agn\_flag & int32 & - &	agn\_flag=0/1: 1 confirmed AGN; 0 AGN candidate \\
8 & sflag & int32 & - &	Method flag: sBL=1,2em=2,sdss=3,else=0 \\
9 & field & bytes10 & - &	Field name in the HETDEX survey \\
10 & nshot & int32 & - &	Number of the observed shots of (ra,dec) \\
11 & nmem & int32 & - &	Number of member detectids \\
12 & detectid\_best & int64 & - &	Detectid closest to the FoF center \\
13 & ra\_best & float32 & deg &	RA of detectid\_best \\
14 & dec\_best & float32 & deg &	DEC of detectid\_best \\
15 & roff & float32 & arcsec &	Offset between ra\_best,dec\_best with fof center \\
16 & mag\_g\_wide & float32 & mag &	HETDEX g measured at (ra\_best, dec\_best) \\
17 & mag\_g\_wide\_err & float32 & mag &	Error of mag\_g\_wide \\
18 & fiberid & bytes38 & - &	ID of the fiber that is closest to detectid\_best \\
19 & shotid & int64 & - &	ID of the shot for detectid\_best \\
20 & multiframe & bytes20 & - &	IFU for detectid\_best \\
21 & throughput & float32 & - &	Throughput of shotid \\
22 & fwhm\_virus & float32 & arcsec &	Image Quality (FWHM) of shotid \\
23 & fwhm\_virus\_err & float32 & arcsec &	Error of fwhm\_virus \\
24 & apcor & float32 & - &	Aperture correction of detectid\_best \\
25 & wobs\_elixer & float32 & Angstrom &	Observed line wavelength fit by ELIXER \\
26 & wobs\_err\_elixer & float32 & Angstrom &	Error of wobs\_elixer \\
27 & flux\_LyA & float32 & 1e-17~\escm &	Rest-frame flux of the LyA emission \\
28 & snr\_LyA & float32 & - &	S/N of LyA \\
29 & fwhm\_LyA & float32 & \kms &	Rest-frame FWHM of the LyA emission \\
30 & ew\_LyA & float32 & Angstrom &	Rest-frame EW of the LyA emission \\
31 & flux\_CIV & float32 & 1e-17~\escm &	Rest-frame flux of the CIV emission \\
32 & snr\_CIV & float32 & - &	S/N of CIV \\
33 & fwhm\_CIV & float32 & \kms &	Rest-frame FWHM of the CIV emission \\
34 & ew\_CIV & float32 & Angstrom &	Rest-frame EW of the CIV emission \\
35 & flux\_CIII & float32 & 1e-17~\escm &	Rest-frame flux of the CIII emission \\
36 & snr\_CIII & float32 & - &	S/N of CIII \\
37 & fwhm\_CIII & float32 & \kms &	Rest-frame FWHM of the CIII emission \\
38 & ew\_CIII & float32 & Angstrom &	Rest-frame EW of the CIII emission \\
39 & flux\_MgII & float32 & 1e-17~\escm &	Rest-frame flux of the MgII emission \\
40 & snr\_MgII & float32 & - &	S/N of MgII \\
41 & fwhm\_MgII & float32 & \kms &	Rest-frame FWHM of the MgII emission \\
42 & ew\_MgII & float32 & Angstrom &	Rest-frame EW of the MgII emission \\
43 & flux\_OVI & float32 & 1e-17~\escm &	Rest-frame flux of the OVI emission \\
44 & snr\_OVI & float32 & - &	S/N of OVI \\
45 & fwhm\_OVI & float32 & \kms &	Rest-frame FWHM of the OVI emission \\
46 & ew\_OVI & float32 & Angstrom &	Rest-frame EW of the OVI emission \\
47 & wave\_pipe & float32 & Angstrom &	Observed line wavelength from the HETDEX pipeline \\
48 & flux\_pipe & float32 & 1e-17~\escm &	Observed line flux from the HETDEX pipeline \\
49 & continuum\_pipe & float32 & 1e-17~\escma &	Observed continuum from the HETDEX pipeline \\
50 & linewidth\_pipe & float32 & Angstrom &	Observed sigma from the pipeline's single Gaussian fit \\
51 & sn\_pipe & float32 & - &	S/N from the pipeline's single Gaussian fit \\
52 & chi2\_pipe & float32 & - &	chi2 from the pipeline's single Gaussian fit \\
53 & chi2fib\_pipe & float32 & - &	chi2 of fit to 2d spec for each fiber \\
54 & rsep\_dr16q & float64 & arcsec &	Separation between the matched SDSS DR16Q AGN \\
55 & z\_dr16q & float32 & - &	Redshift of the matched SDSS DR16Q AGN \\
56 & sdssid\_dr16q & bytes18 & - &	SDSS name of the matched SDSS DR16Q AGN \\ \hline\hline
\label{t_catalog}
\end{longtable}    



\section{Example HETDEX AGN spectra}\label{sec_spec}

\begin{figure}[htbp]
\centering
\includegraphics[width=0.48\textwidth]{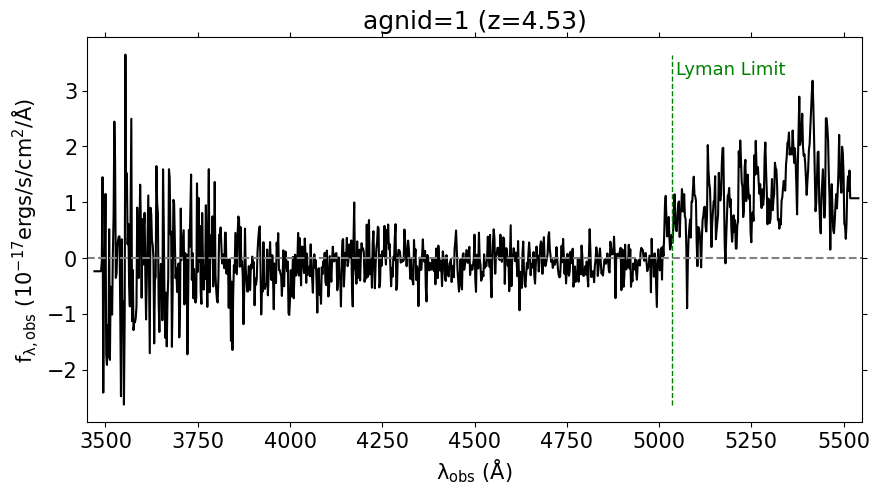}
\includegraphics[width=0.48\textwidth]{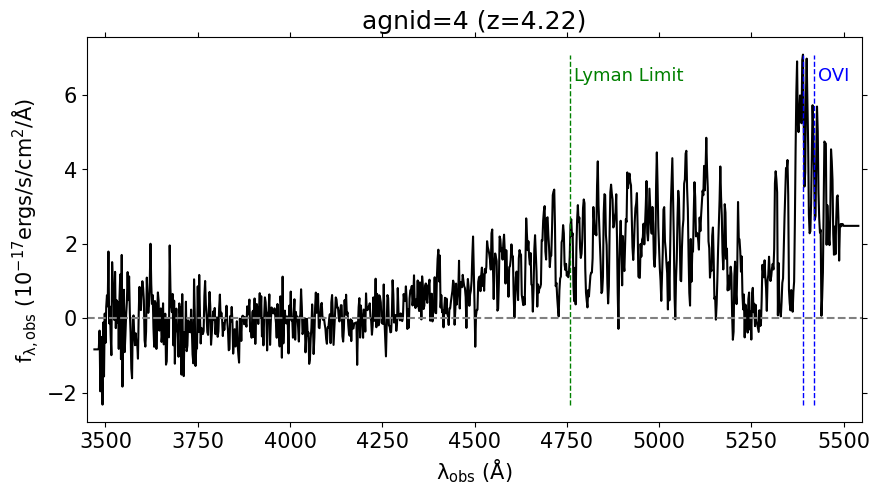}
\includegraphics[width=0.48\textwidth]{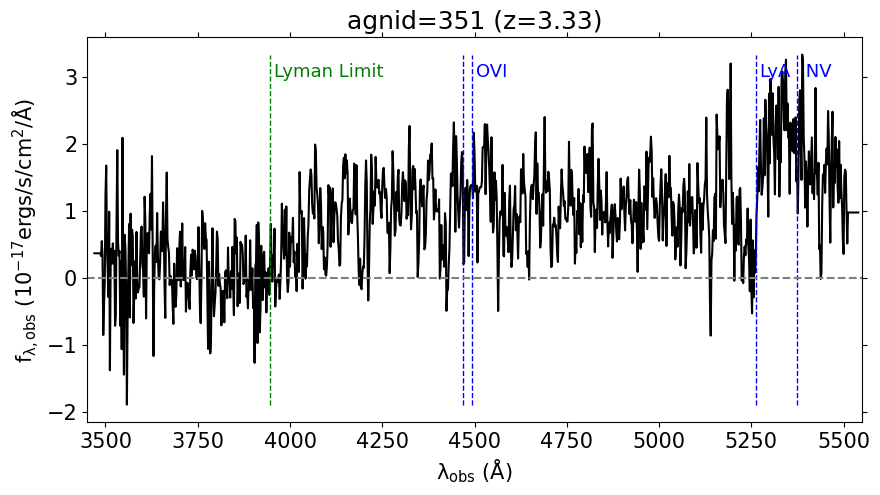}
\includegraphics[width=0.48\textwidth]{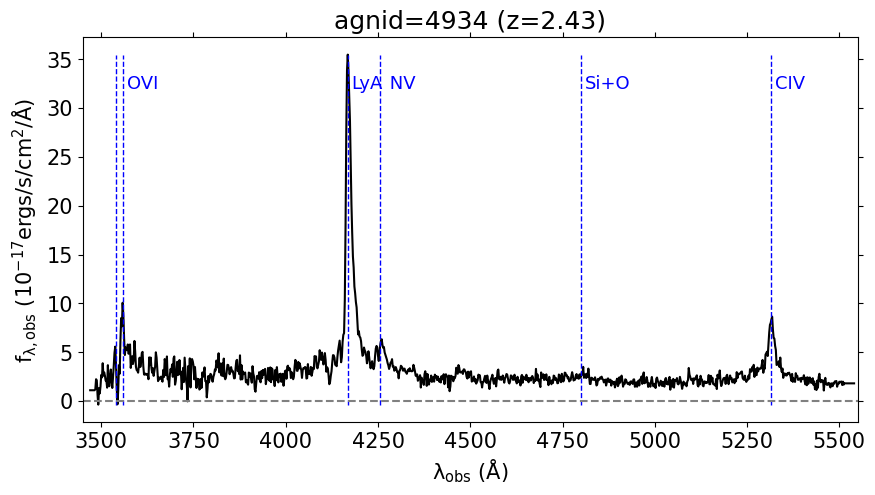}
\includegraphics[width=0.48\textwidth]{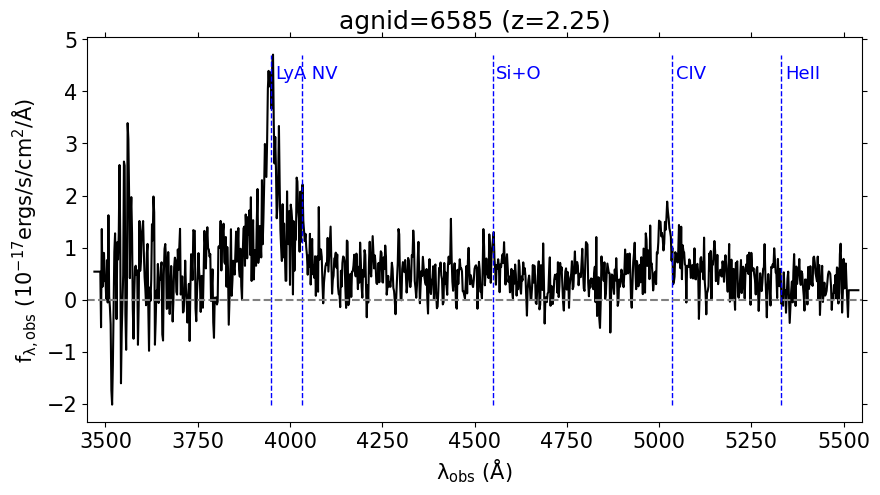}
\includegraphics[width=0.48\textwidth]{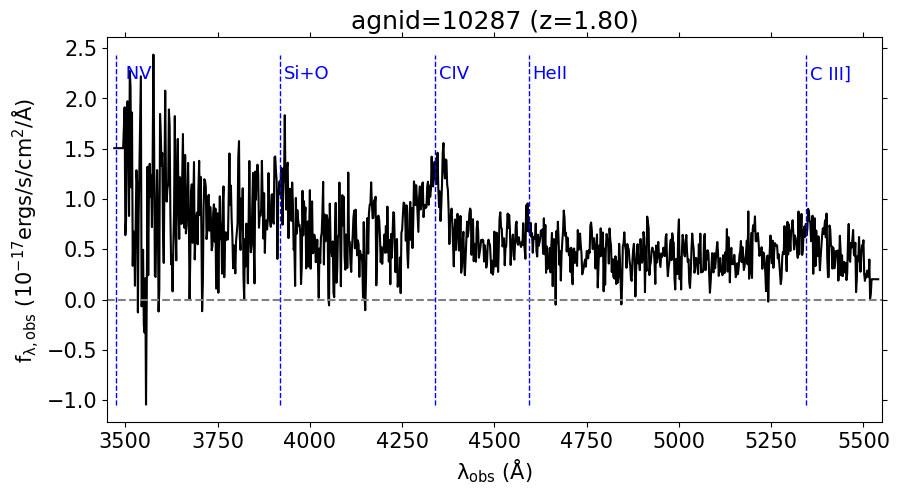}
\includegraphics[width=0.48\textwidth]{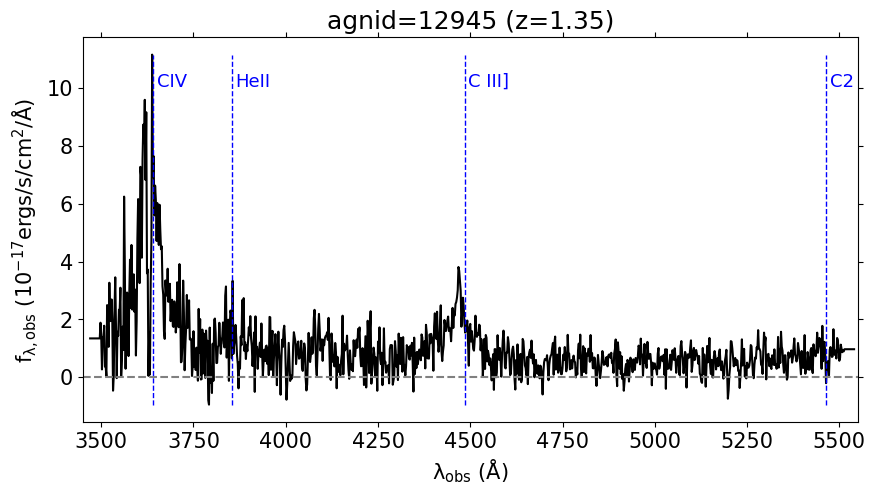}
\includegraphics[width=0.48\textwidth]{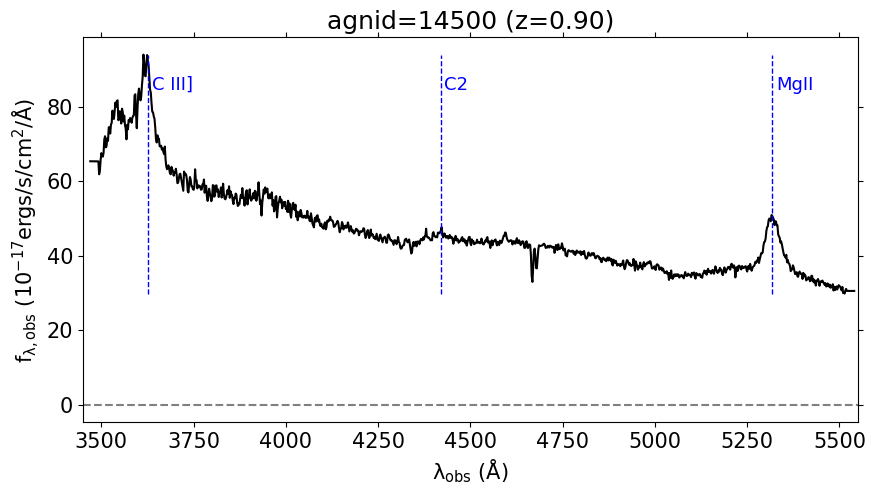}
\includegraphics[width=0.48\textwidth]{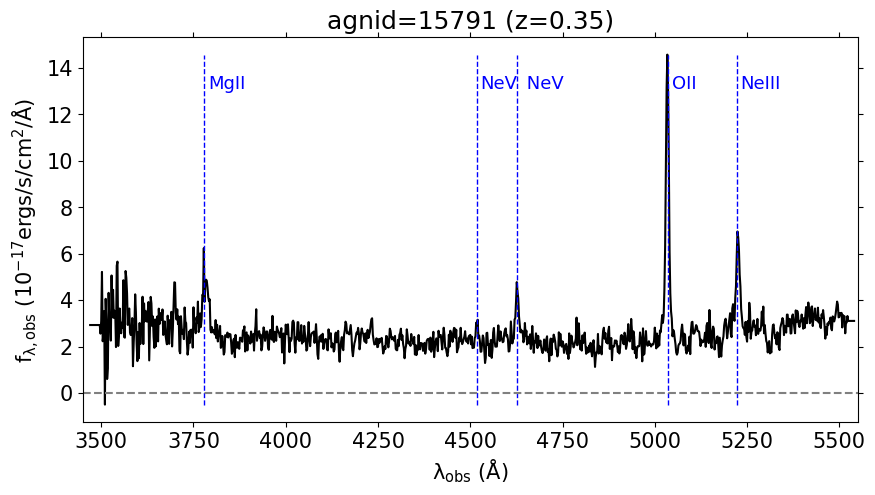}
\includegraphics[width=0.48\textwidth]{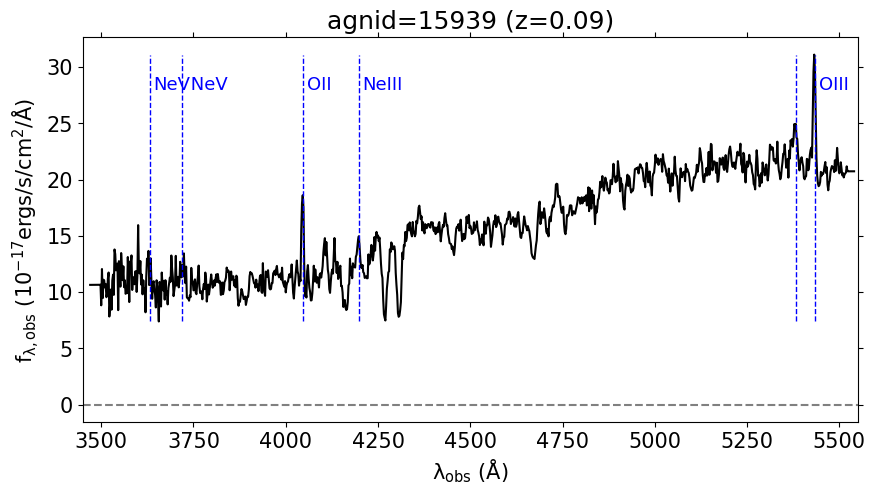}
\caption{Example HETDEX AGN spectra covering the redshift range of the catalog. The resolution of the spectra is 2~\AA.}
\label{f_specs}
\end{figure}

\bibliography{sample631}{}
\bibliographystyle{aasjournal}

\end{document}